\documentclass{aa}  

\usepackage{graphicx}

\usepackage{txfonts}
\usepackage{booktabs}
\usepackage{amsmath}

\begin{document}

   \title{Birthplaces of X-ray emission lines in Cygnus X-3}

   \author{O. Vilhu
          \inst{1}
          \and
          K.I.I. Koljonen
          \inst{2}
          }

    \institute{Dept of Physics, P.O. Box 84, FI-00014  University of Helsinki, Finland\\
              \email{osmi.vilhu@gmail.com}
         \and
             Institutt for Fysikk, Norwegian University of Science and Technology, H\"{o}gskoleringen 5, Trondheim, 7491  Norway\\
             \email{karri.koljonen@ntnu.no}
             }

    \date{Received 2.4.2025; accepted 12.6.2025}
 
  \abstract
   {}
   {
   We investigate the formation of X-ray emission lines in the wind of the Wolf-Rayet (WR) companion in Cyg X-3 by analyzing their orbital dynamics using $\it{Chandra}$ High Energy Transmission Grating (HEG) observations during a hypersoft state. Our goal is to constrain the X-ray transparency of the recently discovered funnel-like structure surrounding the compact star, as revealed by X-ray polarimetry. 
   }
   {
   We analysed $\it{Chandra}$/HEG observations and measured radial velocities and emission line intensities as a function of orbital phase for six emission lines: \ion{Si}{xiv} 6.184 $\AA$, \ion{S}{xvi} 4.734 $\AA$, \ion{Ar}{xviii} 3.731 $\AA$, \ion{Ca}{xx} 3.018 $\AA$, \ion{Fe}{xxv} 1.859 $\AA$, and \ion{Fe}{xxvi} 1.778 $\AA$. We constructed radial velocity and emission intensity light curves for these lines using ten phase bins, which allowed us to investigate their orbital variability using a simple spherical WR-wind model.  
}
   {
   All lines exhibit sinusoidal orbital modulation, with the velocity amplitude generally increasing and the orbital phase with the highest blueshift generally decreasing for ions with a higher ionisation potential. The \ion{Fe}{xxvi}-line displays velocity extremes at phase 0.25 (blueshift) and 0.75 (redshift), with a velocity amplitude of $\sim$500 km/s, indicating that the line-emitting region is close to the compact component (disc or corona) and thus reflects the orbital motion. The \ion{Fe}{xxv}-line shows a complex behaviour that cannot be fully resolved with the $\it{Chandra}$/HEG resolution. Other lines display velocity extremes scattered around phase 0.5 (blueshift) and 0.0 (redshift), with velocity amplitudes of 100--300 km/s, suggesting their origin in the WR stellar wind between the two components.
   }
   {The \ion{Fe}{xxvi}-line originates in the disc or corona of the compact object and can be used to constrain the system masses (Appendix A). The origin of the \ion{Fe}{xxv}-line remains uncertain due to the limitations of $\it{Chandra}$/HEG resolution. The other lines formed in the WR-wind between the component stars, likely above the orbital plane along ionisation $\xi$-surfaces. Parts of the emission lines of \ion{Ar}{xviii} and \ion{Ca}{xx} originated around the compact star. The wind-related component of the \ion{Ca}{xx} line formed closest to the ionizing source (the compact star) due to its highest ionisation potential. The recent polarisation funnel-modelling is consistent with the present results during the hypersoft state.
   } 


   \maketitle

\section{Introduction}

Cygnus X-3 (Cyg X-3; 4U 2030$+$40), one of the first X-ray binaries (XRBs) discovered \citep{giacconi67}, is among the brightest objects in both X-ray \citep[e.g.][]{hjalmarsdotter08,szostek08} and radio wavelengths \citep[e.g.][]{waltman94,waltman96}. It is a high-mass X-ray binary with a donor star (optical counterpart V1521 Cyg) that is likely a Wolf-Rayet (WR) star with a powerful radiatively driven wind, classified as either a WN 5-7 type star \citep{vankerkwijk96} or a weak-lined WN 4-6 \citep{koljonen17}. The relatively small size of this helium star allows for a tight orbit with a period of 4.8 hours \citep{parsignault72}. This characteristic is more typical of low-mass X-ray binaries, thus making Cyg X-3 a unique source in the Galaxy \citep{lommen05}. Recently, \citet{veledina24} predicted the presence of a funnel around the X-ray source based on X-ray polarimetry. This funnel may influence the irradiation of the WR star and its surrounding wind as well as the observed X-ray continuum.

{In this paper, we reanalyse $\it{Chandra}$ High Energy Transmission Grating (HETG) observations taken during a hypersoft state \citep{koljonen10,koljonen18} in January 2006 \citep[initially analysed in][ hereafter Paper I]{vilhu09}. By studying the radial velocities and light curves of multiple emission lines, we aim to determine their formation sites and assess whether the funnel either leaks or channels X-ray flux from the compact object. We infer the formation sites by comparing the K-amplitudes and orbital phases with the highest blueshift of the emission lines with predictions from a spherical WR-wind model. Compared to Paper I,  we employ a more refined continuum treatment, include additional emission lines, and address a new objective: identifying the birthplaces of X-ray emission features. }

\section{Chandra observations}    

\begin{figure}
   \centering
 \includegraphics[width=9cm]{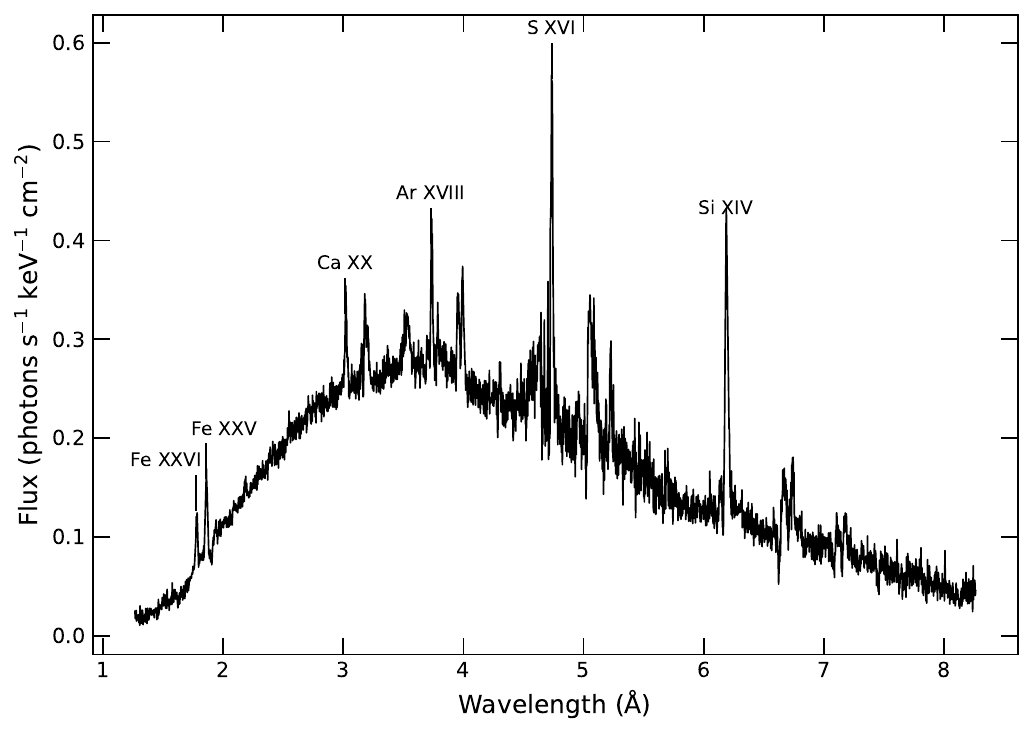}
     \caption{$\it{Chandra}$/HEG spectrum (OBSID 7268). The lines we analysed are marked. {The flux (F) is in units of photons/$s/keV/cm^2$ at Earth.}}
         \label{fig1}
   \end{figure}

$\it{Chandra}$ HETG observations were conducted on January 25–26, 2006, as part of a large international campaign to study Cyg X-3 during an active state (OBSID: 7268, PI: M. McCollough; see Paper I). The observations began on MJD 53760.59458 (corresponding to orbital phase 0.053) and continued until MJD 53761.42972 (orbital phase 4.240). During this period, Cyg X-3 was in a high-soft state (also known as the hypersoft state) with an average $\it{RXTE}$/ASM count rate of approximately 30 cps, equivalent to 400 mCrab. For comparison, typical hard-state count rates are below 10 cps.

We analysed data from the High Energy Grating (HEG), focusing on first-order spectra. The observation spanned more than four orbital cycles and was divided into ten phase bins. We concentrated on six emission lines: \ion{Si}{xiv} 6.184 $\AA$, \ion{S}{xvi} 4.734 $\AA$, \ion{Ar}{xviii} 3.731 $\AA$, \ion{Ca}{xx} 3.018 $\AA$, \ion{Fe}{xxv} 1.859 $\AA$, and \ion{Fe}{xxvi} 1.778 $\AA$. These are not blends, and we could identify them using XSTAR.\footnote{https://heasarc.gsfc.nasa.gov/docs/software/xstar/xstar.html} Nearly all of these lines are hydrogen-like (H-like, with a single remaining electron) except \ion{Fe}{xxv}, which is helium-like (He-like, with two remaining electrons). The spectral resolution was 0.012 $\AA$, corresponding to velocity resolutions of 2020 km/s for the iron lines and 580 km/s for the \ion{Si}{xiv} line. {The lines are marked in Fig.~\ref{fig1}, which shows the calibrated total flux at Earth (photons$/s/keV/cm^2$).  Mean emission line fluxes are given in Table 1, first row. The HEG first-order at 2 keV has an effective area around 100 $cm^2$. Hence, on average one  \ion{Si}{xiv}-line photon was registered every 3.6 seconds.}

These observations were also part of an extensive study on photoionisation emission models of Cyg X-3 \citep{kallman19}. The components of the \ion{Fe}{xxv} line (resonance, intercombination, and forbidden) were recently analysed by \citet{suryanarayanan24} using a dataset similar to ours.
 
\section{Orbital modulation of the emission lines}
{ 
The emission lines can be approximated by single Gaussian profiles, except for the \ion{Si}{xiv} line, which exhibits a clear P Cygni profile (Paper I). The sensitivity and resolution of $\it{Chandra}$ (HEG) does not allow for the absorption components of other lines to be resolved. Thus, single Gaussian fits were used in this study. However, we also include the results obtained from modelling the \ion{Si}{xiv} line with a P Cygni profile ( Figs.~\ref{fig3} -~\ref{fig5a} ). The spectra (photons/$s/keV/cm^2$, see Fig.B.1) from ten orbital phase bins were transformed into velocity, V(km/s), coordinates relative to the rest-frame wavelength of each line ($V_{\rm line}$ km/s) and approximated using the following equation:
}
\begin{equation}
F(V) = F_{\rm cont}(V) + A_{0} \exp\Bigg(\frac{-Z(V)^2}{2}\Bigg),
\end{equation} 
 {
\noindent where $F_{\rm cont}(V) = A_{3} + A_{4} (V - V_{\rm line})$ and $Z(V) = (V - V_{\rm line} - A_{1})/A_{2}$. The five parameters  $A_{0}$ (photons/$s/keV/cm^2$),  $A_{1}$ (km/s), $A_{2}$ (km/s), $A_{3}$ (photons/$s/keV/cm^2$), and $A_{4}$ (photons/$s/keV/cm^2/(km/s)$) were obtained by fitting the observed spectra using the IDL procedure \textsc{mpcurvefit.pro}.\footnote{Written by Craig Markward} We note that the continuum at the line positions was included in the fits. The fitting  was restricted to range between $-$10000 km/s and 10000 km/s. The line widths -- full width at half maximum FWHM $ = 1.38 A_{2}$ -- are primarily instrumental (resolution-limited) and do not provide additional physical information. The fitted line profiles are shown in Fig. B.1.}

{The radial velocities, $V_{r}$ ($= A_{1}$ km/s), at ten orbital phases were fitted with a sine function:}

\begin{equation}
V_r = B_{0} + B_{1} \sin(2\pi\phi + B_{2}).
\label{sini}
\end{equation}
{
\noindent Here, $\phi$ is the orbital phase. This fit yielded the K-amplitude ($= B_{1}$ km/s), the systemic velocity ($= B_{0}$ = $\Gamma$ km/s), and the phase at which the sine curve reaches its extrema ($ B_{2}$ = $\pi$ and  $\pi$/2 for maximum blueshift at phases 0.25 and 0.5, respectively). }

{
The line emission $F_{\rm em}$ (photons/$s/cm^2$) was obtained by integrating  the second term in Eq.1 between -6000 and 6000 km/s. 
We modelled this orbitally modulated total line emission as }

\begin{equation}
F_{\rm em} = C \exp(-A_{\rm wind}\tau_{\rm wind} - A_{\rm clump}\tau_{\rm clump} ),
\end{equation}

\noindent where we assumed that the line flux is absorbed by the same two components: the stellar wind and the so-called clumpy trail, which has been suggested to explain the orbital modulation of the X-ray and infrared continuum emission \citep{vilhu13,antokhin22}.
{
We adopted the same phase-dependent profiles for dimensionless $\tau_{wind}$ and $\tau_{clump}$ as in \citet[][ their Fig. 5]{vilhu13}. The free parameters $C$ (photons/$s/cm^2$), dimensionless $A_{wind}$, and $A_{clump}$ were then determined using \textsc{mpcurvefit.pro}. We also applied a similar fit to the X-ray continuum (co-adding $F_{\rm cont}$ between -6000 and 6000 km/s). }

{
 The fitting results are summarised in Table 1, with the last  row providing chi-squared values for the radial velocity fits (DOF = 7). The value of $F_{em}$ for the  six lines , averaged over the orbital phase, are given in Table 1, first row. Some values of Table 1 can be compared with those reported  in Paper I , which gives  the K-amplitude  and bluest phase for the \ion{Fe}{xxvi}-line: 454$\pm{108}$ km/s and 0.23$\pm{0.04}$, respectively. The corresponding values for the \ion{S}{xvi}-line are  100$\pm{36}$ and 0.44$\pm{0.08}$. These values are consistent, within uncertainties,  with those presented in Table 1. Minor differences can arise from different treatment of the continuum, which in the present paper is improved. }

We note that the physical nature of the clumpy trail remains highly uncertain. It has been suggested to explain the observed decrease in the orbital modulation of the X-ray continuum at $\phi\sim0.4$ \citep{vilhu13}. In this scenario, the (conical) jet emission triggers the formation of dense clumps in the stellar wind, which are then advected along the wind to form a trail producing absorption at a narrow orbital phase interval. High-density clumps may prevent efficient cooling and enhance line absorption. The clumpy trail region is located at approximately 1.4 binary separations from the system’s centre and is crossed by the line of sight to the compact object at orbital phases between 0.2 and 0.6. 

\begin{figure}
   \centering
 \includegraphics[width=9cm,height=16cm]{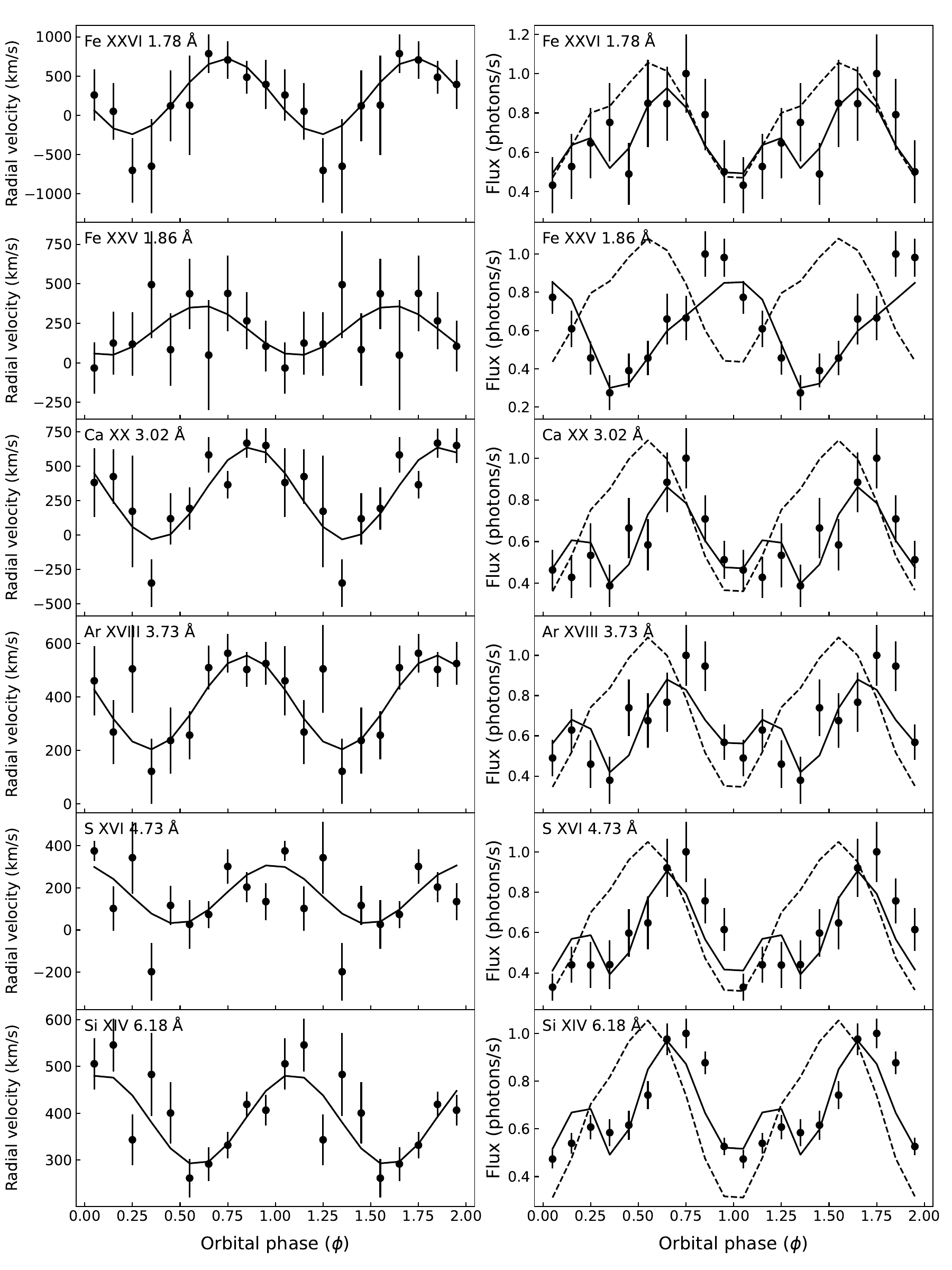}
     \caption{{Sine curve fits for the radial velocity  (km/s, left panels) and two-absorber fits for the line emission (scaled, photons/s, right panels) for the six analysed emission lines. The dashed line represents the continuum at the line.}}
     \label{fig2}
   \end{figure}

\begin{figure}
   \centering
 \includegraphics[width=9cm,height=10cm]{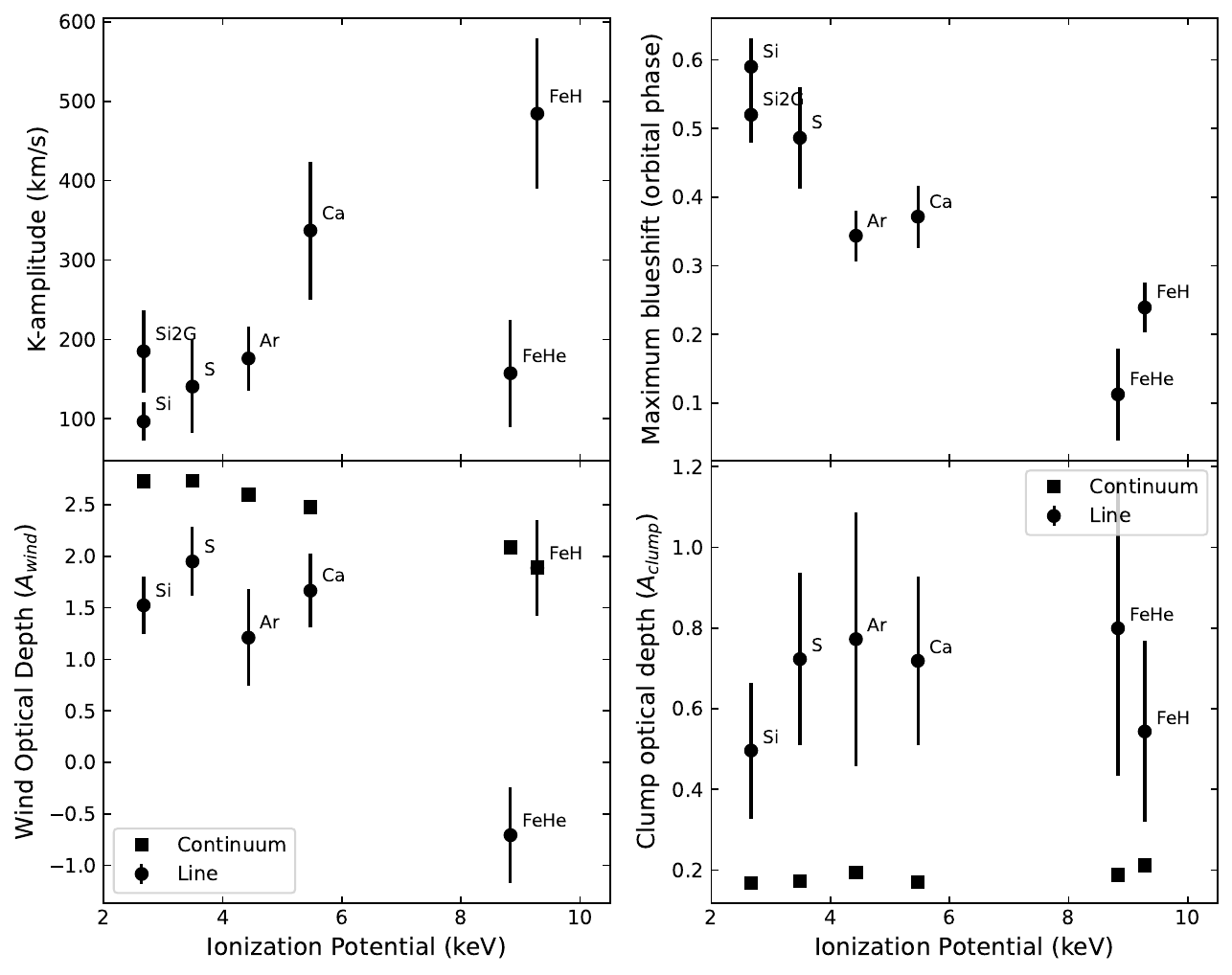}
     \caption{{K-amplitude (upper-left), the orbital phase with the highest blueshift (upper-right), $A_{wind}$ (lower left), and $A_{clump}$ (lower right) as a function of the ionisation potential   of the ion. Dots with error bars show the emission lines, while the filled boxes indicate the continua at the lines. 'Si2G' in the upper panels denotes the P Cygni (two-Gaussian) solution for the \ion{Si}{xiv} line.}}
     \label{fig3}
   \end{figure}

\begin{figure}
   \centering
\includegraphics[width=9.0cm, height=21cm]{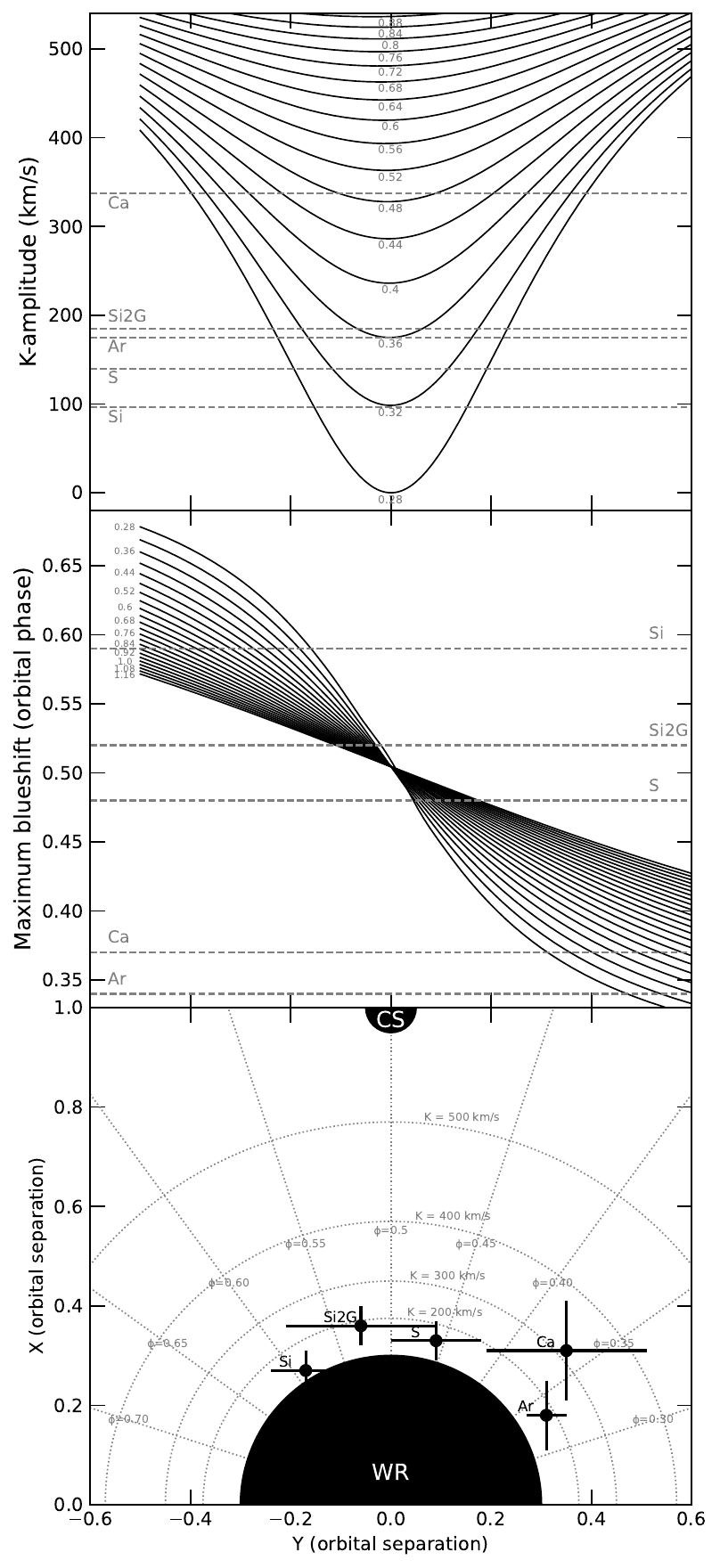}
     \caption{{\textit{Top panel}: K-amplitudes of grid points  plotted against their y-values sorted by their x-values. \textit{Middle panel}: Same as the upper panel but for the orbital phases with the highest blueshift. The values corresponding to the observed emission lines are marked by horizontal lines. \textit{Bottom panel}: Line formation sites  combining information from the two upper panels (dots with error  bars). The bottom panel also shows (dotted) curved contours with constant K-values (200, 300, 400, and 500 km/s increasing from the WR surface) and radial (dotted) contours with the constant
bluest phases $\phi$ (from 0.3 to 0.7 ). } }

     \label{fig4}
   \end{figure}
   
Figure~\ref{fig2} presents the radial velocity (left panels) and light curve fits (scaled by maximum value, right panels) for the six lines studied here. In the right-hand side panels, the continuum fits at the line positions are shown as dashed curves. We note that the rather unclear orbital modulation seen for the radial velocity of \ion{Fe}{xxv} may stem from not resolving the absorption and emission components that are not in phase, as found recently by \citet{xrism24}.

In Fig.~\ref{fig3} (upper panels), we plot the line ionisation potential (which correlates with their photon energy; see Table 1) with the K-amplitude  and the orbital phase with the highest blueshift. Both show  a relationship with the ionisation potential. As the ionisation potential increases, the K-amplitude increases from $\sim$100 km/s to $\sim$500 km/s, while the orbital phase with the highest blueshift decreases from $\phi=0.6$ to $\phi=0.25$. The \ion{Fe}{xxvi}-line emitting region likely {resides in the accretion disc or corona around the compact object}, with the highest blueshift occurring at the orbital phase when the compact object is at its ascending node.
Lower ionisation potential lines with modest K-amplitudes of 100--200 km/s and the orbital phases with the highest blueshift occurring at phases  $\phi=0.4-0.6$ are more consistent in being located closer to the WR star and the centre of mass of the binary, where the stellar wind is  still accelerating and the wind velocities are modest. The \ion{Ca}{xx} line has a higher K-amplitude, indicating formation in a region with a higher wind velocity than the \ion{Ar}, \ion{S}, and \ion{Si} lines.  

In Fig.~\ref{fig3} (lower panels), we plot the line ionisation potential with the optical depths of the wind and clump absorbers. The clump absorption is significantly higher for the lines than for their corresponding continua (shown by filled boxes), while the opposite is true for wind absorption. The strong absorption by the clumpy trail (compared to continuum absorption) suggests that the gas in this region is highly ionised, enhancing line absorption. The \ion{Fe}{xxvi} line  exhibits the same wind absorption as its continuum, indicating a common formation site (compact star). The other emission lines likely originate above the orbital plane, as their wind absorption is lower than that of the continuum (see Fig.~\ref{fig3}, lower-left plot). 

Finally, we note that the continuum wind absorption depends on wavelength (or ionisation potential), as shown in Fig.~\ref{fig3} (lower-left plot). This behaviour resembles the differences observed between soft and hard X-ray light curves \citep{vilhu03,zdziarski12}. Pure wavelength-independent Thomson scattering alone cannot fully explain this trend.

\begin{table*}
    \centering
    \caption{Results of the fitting procedures using a sine-curve for radial velocities and two absorbers for line intensities.}
    \label{tab:fitting_results}
    \renewcommand{\arraystretch}{1.2}
    \begin{tabular}{lcccccc}
        \toprule
        & \ion{Fe}{xxvi} & \ion{Fe}{xxv} & \ion{Ca}{xx} & \ion{Ar}{xviii} & \ion{S}{xvi} & \ion{Si}{xiv}  \\
        & (1.778 \AA) & (1.859 \AA) & (3.018 \AA) & (3.731 \AA) & (4.734 \AA) & (6.184 \AA) \\
        \midrule
$F_{\rm em}$ ($10^{-3}$  photons/$s/cm^2$)  &3.8 &5.2 & 2.1 & 1.9 &3.5 &2.7  \\
        $E_{\rm line}$ (keV)         & 6.97  & 6.67  & 4.11  & 3.32  & 2.62  & 2.00  \\
        $E_{\rm ion}$ (keV) & 9.28  & 8.83  & 5.47  & 4.43  & 3.49  & 2.67  \\
        \midrule
        $K$ (km/s)                  & $484 \pm 95$  & $157 \pm 67$  & $337 \pm 87$  & $175 \pm 40$  & $140 \pm 58$  & $96 \pm 23$  \\
        $\phi_{\rm max, blueshift}$              & $0.23 \pm 0.03$  & $0.11 \pm 0.06$  & $0.37 \pm 0.04$  & $0.34 \pm 0.03$  & $0.48 \pm 0.07$  & $0.59 \pm 0.04$  \\
        \midrule
        $\Gamma$ (km/s)           & $242 \pm 79$  & $204 \pm 48$  & $301 \pm 69$  & $379 \pm 30$  & $169 \pm 44$  & $386 \pm 19$  \\
        FWHM (km/s)               & $1772 \pm 516$  & $1710 \pm 310$  & $1077 \pm 254$  & $694 \pm 142$  & $618 \pm 129$  & $836 \pm 62$  \\
        \midrule
        $A_{\text{wind}}$ (Line)  & $1.88 \pm 0.46$  & $-0.70 \pm 0.46$  & $1.66 \pm 0.35$  & $1.21 \pm 0.46$  & $1.94 \pm 0.33$  & $1.52 \pm 0.28$  \\
        $A_{\text{clump}}$ (Line) & $0.54 \pm 0.22$  & $0.79 \pm 0.36$  & $0.71 \pm 0.20$  & $0.77 \pm 0.31$  & $0.72 \pm 0.21$  & $0.49 \pm 0.16$  \\
        \midrule
        $A_{\text{wind}}$ (Cont.)  & $1.88 \pm 0.34$  & $2.08 \pm 0.30$  & $2.47 \pm 0.34$  & $2.59 \pm 0.33$  & $2.73 \pm 0.43$  & $2.72 \pm 0.41$  \\
        $A_{\text{clump}}$ (Cont.) & $0.21 \pm 0.13$  & $0.18 \pm 0.13$  & $0.17 \pm 0.14$  & $0.19 \pm 0.14$  & $0.17 \pm 0.17$  & $0.16 \pm 0.17$  \\
        \midrule
        $\chi^2$   (DOF=7)       & 0.19  & 0.35  & 0.24  & 0.08  & 0.28  & 0.03  \\
        \bottomrule
    \end{tabular}
\end{table*}
\section{Sites of line formation}

In this section, we study the potential locations for the line emission in more detail. To obtain a rough estimate, we created a rotating grid of mesh points in the orbital (x,y) plane between the stars and computed radial velocities within this grid. 
{In this coordinate system, the x-axis connects the stars, the y-axis is perpendicular to it, the WR  star  is  located at (x,y) = (0,0), and the compact object is at  (1,0). The coordinate system rotates counterclockwise around the WR centre over the  4.8-hour orbital period.  The grid consists of $80  \times 80$  points spanning from  (x,y) = (0,-0.5) with a uniform step size of  $\Delta$x = $\Delta$y = 0.014.}

The wind velocity was modelled using the $\beta$-law: 

\begin{equation}
 v_{\rm wind} = v_{\infty}(1 - R_{star}/r)^{\beta},
\end{equation}

where $v_{\infty} = 1600$ km/s and $\beta = 1.0$, which are values consistent with WNE stars in the Potsdam models.\footnote{\url{http://www.astro.physik.uni-potsdam.de/~wrh/PoWR/WNE/}} We adopted an orbital inclination of $i=30^\circ$ \citep[][see also Appendix A]{vilhu09,antokhin22,veledina24a}.

{
At each grid point, we computed the K-amplitude and the orbital phase of the highest blueshift, then compared these values with the observed parameters 
of the six emission lines listed in Table 1. This analysis is illustrated in Fig.4.
In the top panel of Fig.4, the K-amplitudes of the grid points are plotted as a function of their y-values. Curves are labelled by the corresponding x-coordinate. 
The middle panel shows the corresponding orbital phases of the highest blueshift. Observed values from Table 1 are indicated in both panels by dashed horizontal 
lines. In the figure, `Si2G' refers to a two-Gaussian fit (emission plus absorption, i.e. P Cygni profile). The bottom panel summarises the inferred formation regions by 
combining information from the top two panels, including error estimates derived from the data. Dotted contours indicate constant K-amplitudes and orbital phases with highest blueshift.
 In practice, the likely formation site of each emission line corresponds to the intersection of contours representing its observed K-amplitude and bluest phase, 
as listed in Table 1.
}

 The plot in the bottom panel of  Fig.~\ref{fig4} depicts the system at phase $\phi = 0$ as viewed from below. The plot is on the orbital plane (z = 0). However, the birth sites are briefly the same when using z = $\pm{0.3}$.
The birthplaces of the \ion{Ca}{xx} and \ion{Ar}{xviii} lines are not symmetrically located between the stars because their orbital phase with the highest blueshift clearly occurs
before phase 0.5. This may indicate that a part of them formed near the compact star, which we examine in the next section.

\begin{figure}
   \centering
 \includegraphics[width=8cm]{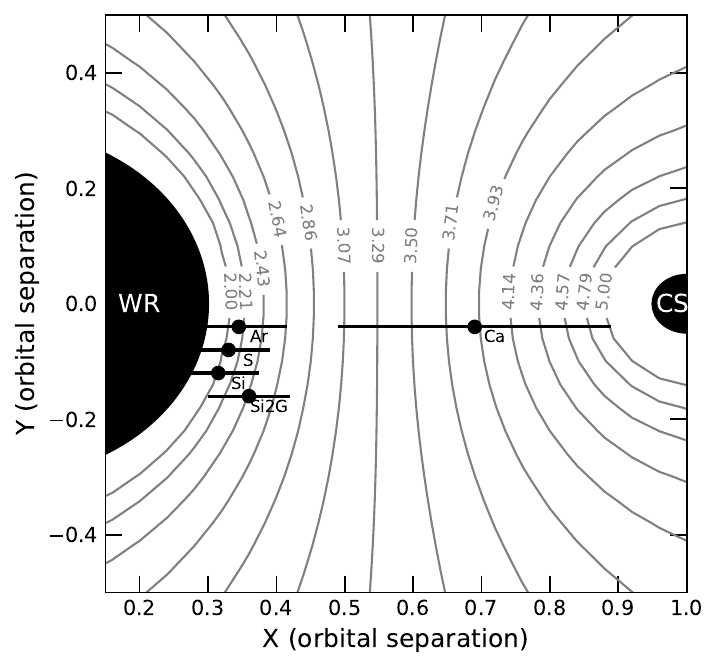}
     \caption{{Alternative arrangement for the emission regions of the \ion{Si}{xiv}, \ion{S}{xvi}, \ion{Ar}{xviii}, and \ion{Ca}{xx} lines. Parts of the Ca and Ar lines formed in the compact star disc or corona (see the text). The solid curves represent surfaces with a constant ionisation parameter marked by a log $\xi$-value. These curves are intersections with the (x,y) plane of  3D surfaces where the lines are expected to form.  For clarity, the error bars have been shifted vertically from y=0. The coordinate system is the same as in Fig. \ref{fig4}.}}
     \label{fig5a}
   \end{figure}
We assumed that the Ca and Ar lines originate from two locations: the stellar wind, where the maximum blueshift occurs at phase $\phi = 0.5$, and another region near the compact star, where the maximum blueshift occurs at phase $\phi = 0.25$. Additionally, we assumed that the \ion{Fe}{xxvi} line formed entirely in the vicinity of the compact star (disc or corona). This is also supported by XSTAR wind-simulations, which do not provide enough  \ion{Ar}{xviii},  \ion{Ca}{xx}, and  \ion{Fe}{xxvi} line photons (see Fig.2 in Paper I). The observed radial velocity curve can then be expressed as the sum of the contributions from line emission at both the wind and compact star locations:

\begin{equation}
\begin{split}
K & \sin{(2\pi\phi + \phi_{\rm max, blueshift})} = \\
& f K_{\rm wind} \sin(2\pi \phi + \pi/2) + K_{\rm CS} \times (1 - f) \sin(2\pi \phi + \pi).
\end{split}
\end{equation}

\noindent Here, $f$ is the fraction of the line emission attributed to the wind location, while the values for $K$ and $\phi_{\rm max, blueshift}$ can be found in Table 1 for each line. We used $K_{\rm CS} = K_{\rm Fe XXVI}$ = 484 km/s (see Table 1) and solved for $f$ and $K_{\rm wind}$. In the case of \ion{Ca}{xx}, we obtained $K_{\rm wind} = 470$ km/s and $f = 0.5$, indicating that half of the line photons formed in the wind in a region with a relatively high velocity and the other half  formed at the compact star. For \ion{Ar}{xviii}, we obtained $K_{\rm wind} = 140$ km/s and $f = 0.7$, indicating that the majority of the line emission is consistent with arising from the wind close to the WR companion.     

The alternative line formation sites in the stellar wind consistent with the derived parameters are illustrated in Fig.~\ref{fig5a}. The K-values of the S and Si lines were obtained by fixing their bluest phases to $\phi = 0.5$. These values are not significantly different from those in Table 1.

Errors were propagated from uncertainties in K-amplitudes and $\phi_{\rm max, blueshift}$. We also plot the constant ionisation parameter surfaces, $\xi$ = $L_X/(Nr_{CS}^2)$ in the figure while assuming a spherically symmetric X-ray emission from the compact star. Here, $L$ is the ionizing luminosity, estimated as $L = 2.46 \times 10^{38}$ erg/s (Paper I); $N$ is the gas number density; and $r_{CS}$ is the distance from the compact star. In mass-conserving, spherical, and isotropic wind, the mass-loss rate is given by $\dot{M} = 4\pi r^2 \rho v_{\rm wind}$, where the wind density depends on the distance from the WR centre as $\rho \propto r_{WR}^{-2}/v_{\rm wind}$. We computed $ v_{\rm wind}$ using Eq. 4.

Assuming a wind mass-loss rate of $\dot{M}$ = $10^{-5}$ $M_{\sun}$/year \citep{koljonen17} and a mean molecular weight of $\mu = 1.33$ for ionised helium gas, we could derive the constant ionisation surfaces in Fig. 5  in the range $\log{\xi} = 2.0-5.0$, which covers the equilibrium values for our observed lines (\ion{Si}{xiv}: $\log{\xi} = 2.0-3.5$, \ion{Ca}{xx}: $\log{\xi} = 2.5-4.5$, as given in the XSTAR manual).
{
Assuming a wind consisting mostly of singly ionised helium ( $\mu = 1.75$), the above ranges would change only modestly (increased by 0.12).
 In addition, we expect that the HeII-ions are formed at the far side of the WR wind with respect to the compact star (in the X-ray shadow); this is supported by infrared 
observations from Koljonen and Maccarone (2017, their Fig.8). Using the methods described in this paper, the inferred formation sites of the HeII lines lie 
near (x,y) = (–0.55,0) based on the K-amplitude of 400 km/s and orbital phase 0.0  at the highest blueshift.}

{The wind on the face-on side is significantly influenced by X-rays from the compact star. In particular, the EUV flux below the HeII ionisation edge
 at 54.4 eV (228 Å) plays a critical role in regulating the ionisation of species responsible for the radiative line driving 
(primarily NiV, FeV, CIV, NV, and CrV; Vilhu et al. 2021, their Appendix B). When the EUV flux is strong, these ions become fully ionised, reducing
 or eliminating the line-driving force.}

{During the hypersoft state of Cyg X-3, this ionisation occurs roughly halfway between the WR and the compact star, creating a kink in the wind velocity 
profile \citep{vilhu23}. We note that the wind velocities in Fig.2 of that study were computed without including the gravitational pull of the compact
 star in order to facilitate accretion modelling, so they only reflect the impact of the reduced line force. Including the gravitational attraction of the compact star
 would (over)compensate the loss in line force. Nevertheless, this kind of wind suppression model does not alter the qualitative conclusions of the
 present model, though it may shift the inferred \ion{Ca}{xx} line formation sites slightly closer to the WR star. }

Using  constant ionisation $\xi$-surfaces, we can investigate the K-amplitude values over the entire surface, weighting local velocities by the wind density. The derived K-amplitude values, with bluest phases at 0.5, do not alter the qualitative conclusions of this paper.

\section{Discussion}

\subsection{Sensitivity of the parameter values on the emitting locations}

In this section, {we discuss relaxing some of the assumptions} used in the above modelling. Concerning the wind velocity law, we also repeated computations with smaller $\beta$ values and found that the only difference was that the emission locations were closer to the WR star. Furthermore, the same effect was observed when using the X-ray suppressed wind model proposed by \citet{vilhu21} (if the gravitational pull of the compact star is included). Similarly, a larger orbital inclination ($>30^{\circ}$) produced the same effect. We did not account for the impact of the WR star’s orbital motion around the centre of mass on wind velocity. However, this effect is expected to be small since the mass ratio is large (see Appendix A) and the centre of mass lies inside the WR companion.

Most likely, the sizes of the actual line emitting regions are much more extended than those shown in Figs.~\ref{fig4} and \ref{fig5a}, which are based on the closest matches between observed K-values and orbital phases of the highest blueshift at individual grid points. These values likely represent an average of many regions. A co-added combination of multiple grid points, weighted by density and photoionisation effects, can also reproduce the observed K-values and orbital phases of the highest blueshift. However, without a detailed photoionisation study (such as that in \citealt{kallman19}), the precise sizes and locations of the emitting regions remain difficult to map.

\subsection{Implications for the system geometry at the compact star}

The emitting regions of the lines could provide constraints on the potential funnel structure around the compact star \citep{veledina24}, as this funnel must produce or leak sufficient ionizing radiation to illuminate the WR wind on the side facing the compact star. Based on Veledina et al.  scenario (see their Fig. 5), this appears to be the case, at least in the hypersoft state, where radiation is primarily scattered (single scattering) off the material along the funnel axis, rather than being dominated by emission reflected from the funnel walls. For X-rays in the keV range, the Klein-Nishina formula predicts scattering in all directions. The observed reduction of X-ray polarisation in the hypersoft state could also arise from the effects of matter ionisation, in which case the funnel becomes partially transparent to the reflected radiation.

We note that the low-hard state X-ray spectra reveal spectral lines similar to those  shown in the hypersoft state \citep[scaled with the continuum;][]{kallman19}. This may already indicate that the funnel walls in the hard state are also at least partially transparent.

\subsection{Issues in determining the parameters for Fe XXV emission region}

As noted above, the birthplace for the \ion{Fe}{xxv} line remains problematic and cannot be reliably determined with the \textit{Chandra}/HEG resolution. Notably, phase-dependent radiative transfer effects and variations in the line components (intermediate, resonance, and forbidden) may obscure any Doppler motion related to the line formation region. 

Observations with the \textit{XRISM}/Resolve spectrometer, which offers unprecedented spectral resolution, have shown that emission and absorption lines can be distinguished  \citep{xrism24}. The emission lines in the 6--9 keV range originate from a region with a medium ionisation parameter (log $\xi \approx 3$) and exhibit radial velocity modulation with $K = 194 \pm 29$ km/s, and they reach the highest blueshift at the orbital phase $\phi=0.25$ (though the \ion{Fe}{xxv} line should be separated from the \ion{Fe}{xxvi} line). The absorption lines show a much weaker orbital modulation, with $K = 55 \pm 7$ km/s, but they have a significantly higher systemic velocity ($\Gamma = -534\pm6$ km/s). Their radial velocity extrema occur at orbital phases $\phi=0.1$ (maximum) and $\phi=0.6$ (minimum), suggesting they originate from a larger region with a size scale comparable to the binary orbit.

\section{Conclusions}

We have analysed $\it{Chandra}$/HEG observations of Cyg X-3 during a hypersoft state. We determined radial velocities and emission line intensities as a function of orbital phase for six X-ray emission lines: \ion{Si}{xiv} 6.184 $\AA$, \ion{S}{xvi} 4.734 $\AA$, \ion{Ar}{xviii} 3.731 $\AA$, \ion{Ca}{xx} 3.018 $\AA$, \ion{Fe}{xxv} 1.859 $\AA$, and \ion{Fe}{xxvi} 1.778 $\AA$ (Fig.\ref{fig1}).

Assuming a spherical wind and a two-component absorbing model, we determined the potential emission regions for the studied lines. As in Paper I, we assumed that the iron line \ion{Fe}{xxvi} originates in the compact star disc or corona and that the radial velocity curve reflects its orbital motion. In Appendix A, this is used to constrain system parameters. The origin of the He-type iron line, \ion{Fe}{xxv}, remains unresolved due to the insufficient resolution of $\it{Chandra}$/HEG. 

We find that the other lines likely formed within the WR wind  between the two stellar components, with their radial velocities reflecting wind velocities. Parts of \ion{Ca}{xx} and \ion{Ar}{xviii} emissions originate in the disc or corona of the compact star, with their radial velocities representing a mixture of wind and orbital motions. This is also supported by XSTAR wind-simulations (Paper I, \citet{vilhu09}). The ionisation parameter, $\xi$,  required to achieve a given ionisation state is higher for high-Z elements than for lower-Z elements \citep{kallman19}. Consequently, the wind component of the \ion{Ca}{xx} line emission is  located closer to the compact star than the other lines (Fig.~\ref{fig5a}).

The line-emitting regions are also likely located in the upper half of the (more or less  spherical) ionisation  $\xi$-surfaces, as their wind absorption is weaker than that of the continuum originating from the compact star. Assuming different wind models (e.g. lower $\beta$ values) shifts the line formation sites closer to the WR star. A similar effect occurs if the system inclination is higher than 30$^{\circ}$ (for a discussion on inclination, see Appendix A).

The funnel model around the compact star by \citet{veledina24a} appears to allow wind ionisation at these sites, at least during the hypersoft state. Whether the funnel walls remain transparent during the quiescent hard state remains to be studied.

\begin{acknowledgements}

This project has received funding from the European Research Council (ERC) under the European Union's Horizon 2020 research and innovation programme (grant agreement No. 101002352). {We thank the referee for very useful comments.}
\end{acknowledgements}

\bibliographystyle{aa}
\bibliography{bibliography}

\begin{appendix}

\section{Component masses and inclination}

Assuming that the \ion{Fe}{xxvi} line originates close to the compact star (disc/corona), we can derive the WR mass function from the K-value of $K_{c} = 484\pm95$ km/s (Table 1):

\begin{equation}
 f_{\rm WR} = \frac{K_{c}^3P}{2\pi G} =2.35^{+1.66}_{-1.14}M_{\sun} \equiv \frac{M_{\rm WR}^3 \sin^3{i}}{M_{\rm tot}^2},
\label{eqwr}
\end{equation}

\noindent where $P=17253$ sec is the orbital period \citep{antokhin19}, $G$ is the gravitational constant, $i$ is the orbital inclination (the angle between the line of sight and the orbital plane axis), and $M_{\rm WR}$ and $M_{\rm tot}$ are the WR and total system masses, respectively. This value is slightly larger than that computed in Paper 1, where a compromise was made in phase binning (see also \citealt{zdziarski13}).  

\citet{hanson00} observed infrared \ion{He}{i} absorption lines with a K-amplitude of $K_{\rm WR} = 109 \pm 13$ km/s. They favour the interpretation that these lines are centred on or associated with the motion of the WR-star (however, see discussion in \citealt{koljonen17} with possible caveats on the detection of this line). Using this K-value, the mass function of the compact star can be derived as

\begin{equation}
 f_{\rm c} = \frac{K_{\rm WR}^3P}{2\pi G} =0.027^{+0.011}_ {-0.008}M_{\sun} \equiv \frac{M_{\rm c}^3 \sin^3{i}}{M_{\rm tot}^2},
\label{eqc}
\end{equation}

\noindent where $M_{\rm c}$ is the compact star mass. 

Dividing these two mass functions gives the mass ratio

\begin{equation}
 \frac{M_{c}}{M_{\rm WR}} = \frac{K_{\rm WR}}{K_c}= 0.23^{+0.06}_{-0.04}.
\label{massrat}
\end{equation}
Substituting Eq. \ref{massrat} into Eq. \ref{eqwr} and assuming an inclination of 30 degrees, we obtain the WR mass: $M_{\rm WR} = 28^{+17}_{-12} \, M_{\sun}$. This corresponds to a compact object mass of $M_{\rm c} = 6.3^{+2.5}_{-2.0} \, M_{\sun}$. As with all values determined above, we report the mean and the 16\%--84\% confidence interval of the resulting parameter distribution. The 0.2\%--99.8\% interval (3$\sigma$) is $3.6 \, M_{\sun} < M_{\rm WR} < 96 \, M_{\sun}$ and $1.7 \, M_{\sun} < M_{\rm c} < 15 \, M_{\sun}$.

If Eq. \ref{eqc} cannot be trusted, we can instead use Eq. \ref{eqwr} and an inclination of 30 degrees to determine a limiting minimum WR mass for which the compact star mass is at least that of a neutron star ($M_c \geq 1.4 \, M_{\sun}$): $M_{\rm WR} \geq 21^{+14}_{-9} \, M_{\sun}$. At the 3$\sigma$ level, $M_{\rm WR} > 3.2 \, M_{\sun}$. We note that the total mass of the binary, which can be estimated from the orbital decay and the WR mass-loss rate, is constrained to $M_{\rm tot} \lesssim 20 \, M_{\sun}$ \citep{zdziarski13, koljonen17, antokhin22}.
 
The inclination of 30 degrees is based on the following considerations. Simulating infrared \ion{He}{ii} emission and comparing it with the observations of \citet{hanson00}, Paper 1 estimated an inclination of 30 degrees. This emission originates from the X-ray shadow on the WR star’s far side. \citet{antokhin22} determined an inclination of $29.5 \pm 1.2$ degrees based on a light curve study. \citet{veledina24a} provided inclination limits of 26–28 degrees, assuming that the funnel axis is perpendicular to the orbital plane.

\begin{figure}
   \centering
 \includegraphics[width=9cm]{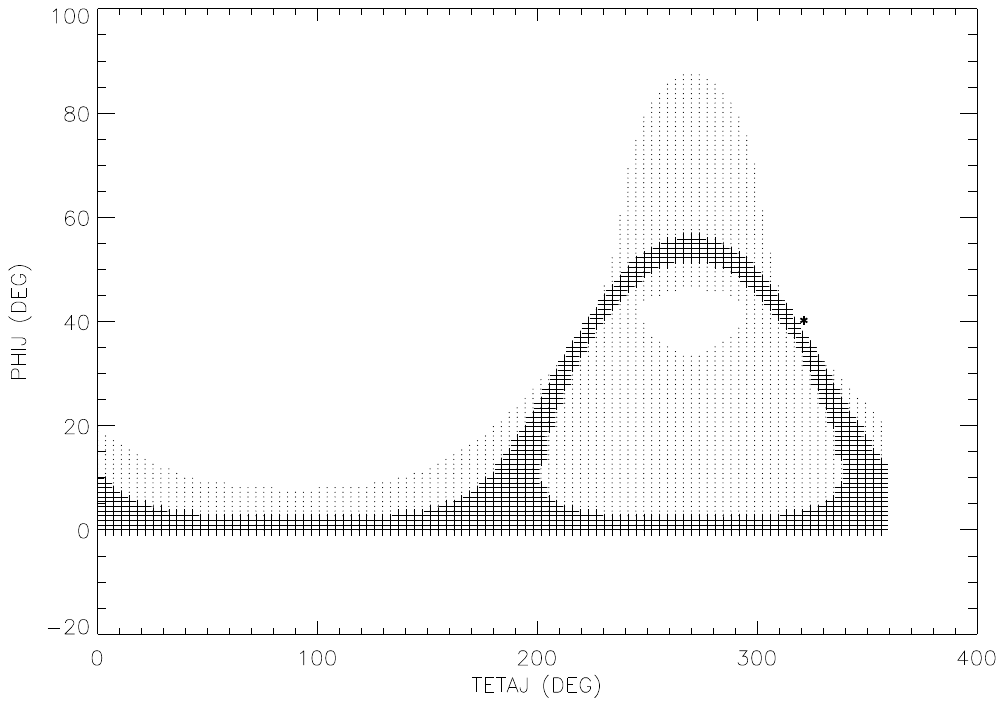}
     \caption{Permitted ($\Theta_j$,$\phi_j$) range corresponding to the angle between the line of sight and the jet (or funnel) axis to lie between 26--28 degrees. The dots indicate constraints on the inclination between 20--60 degrees, while the plus signs (darker region) mark the range between 26--28 degrees. The star represents the jet parameter values given by \citet{dubus10}.}
         \label{fig6}
   \end{figure}

However, there are alternative explanations for the 26-–28 degree angle between the funnel axis and the line of sight. This is demonstrated in Fig.\ref{fig6}, where $\Theta_j$ (TETAJ) represents the jet (or funnel) azimuth (true anomaly) and $\phi_j$ (PHIJ) the polar angle (see Fig. 1 in  \citet{dubus10} ). The ($\Theta_j$, $\phi_j$) values (319, 39) given by \citet{dubus10} closely correspond to a scattering angle of 26-–28 degrees (see the star marker in Fig.\ref{fig6}). These angles define the `clumpy trail' absorber. Fixing the inclination at 30 degrees and adjusting $\Theta_j$ to 317 degrees results in a scattering angle of 27.3 degrees, which falls within the established limits.

\section{Lines observed}  
\begin{figure*}
   \centering
\includegraphics[width=18cm,height=20cm]{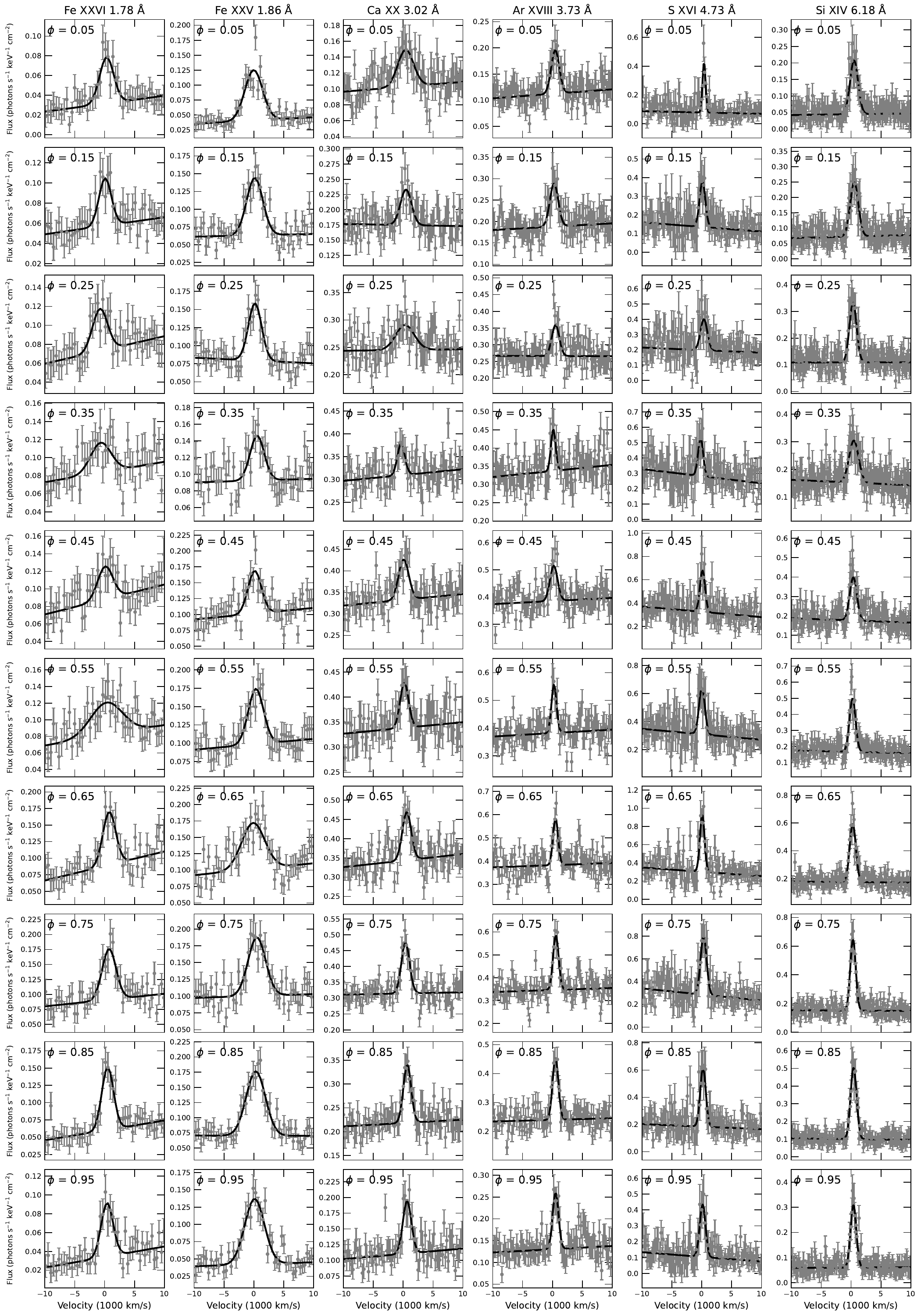}
   \caption{Observed spectra of \ion{Fe}{xxvi} 1.778 \AA, \ion{Fe}{xxv} 1.859 \AA, \ion{Ca}{xx} 3.018 \AA, \ion{Ar}{xviii} 3.731 \AA, \ion{S}{xvi} 4.734 \AA and \ion{Si}{xiv 6.184 \AA}
are shown in vertical columns from left to right. Each column presents spectra at ten orbital phases from 0.05 (top) to 0.95 (bottom). The x-axis represents the velocity shift from the line wavelength (in units of 1000 km/s), while the y-axis shows the observed flux 
(photons/$s/keV/cm^2$ on Earth) with error bars and the model-fit (solid line)}
         \label{fig7}
   \end{figure*}

We collect here all six emission  lines observed at ten phases and present them in   Fig.~\ref{fig7}, which shows the flux  (photons/$s/keV/cm^2$ on Earth)  versus velocity (km/s) from the line centre.

\end{appendix}

\end{document}